\documentclass[manuscript ]{aastex701}

\usepackage{graphicx}
\usepackage{caption}
\usepackage{subcaption} 
 
\begin{document}

\title{The Production Mechanism of the High-energy Emission Line in the Brightest Cosmic Burst}

\author[orcid=0000-0002-7797-9814]{Jieying Liu}
\affiliation{Yunnan Observatories, Chinese Academy of Sciences, Kunming 650216, Yunnan Province, China}
\affiliation{Center for Astronomical Mega-Science, Chinese Academy of Sciences, 20A Datun Road, Chaoyang District, Beijing 100012, China}
 \affiliation{International Centre of Supernovae, Yunnan Key Laboratory, Kunming 650216, P. R. China}
\email[]{ljy0807@ynao.ac.cn}  
 
 \correspondingauthor{Jirong Mao, Shao-Lin Xiong} 
 \author[orcid=0000-0002-7077-7195]{Jirong Mao} 
\affiliation{Yunnan Observatories, Chinese Academy of Sciences, Kunming 650216, Yunnan Province, China}
\affiliation{Center for Astronomical Mega-Science, Chinese Academy of Sciences, 20A Datun Road,
  Chaoyang District, Beijing 100012, China}
 \email[show]{jirongmao@mail.ynao.ac.cn}

\author[orcid=0000-0002-4771-7653]{Shao-Lin Xiong}
\affiliation{State Key Laboratory of Particle Astrophysics, Institute of High Energy Physics, Chinese Academy of Sciences, Beijing 100049, China}
\email[show]{xiongsl@ihep.ac.cn}

\author[orcid=0000-0001-5348-7033]{Yan-Qiu Zhang}
\affiliation{State Key Laboratory of Particle Astrophysics, Institute of High Energy Physics, Chinese Academy of Sciences, Beijing 100049, China}
\affiliation{University of Chinese Academy of Sciences, Chinese Academy of Sciences, Beijing 100049, China}
\email{zhangyanqiu@ihep.ac.cn}

\begin{abstract}
As a characteristic feature of the spectrum, the emission line carries critical information on the underlying physics of the radiation. After extensive efforts in decades, the first high-significant detection of a series of emission lines evolving from 37\,MeV to 6\,MeV has been detected in the ever-bright gamma-ray burst GRB 221009A. However, the physical mechanism of the entire evolutionary trend of the lines remains elusive. To provide a self-consistent interpretation, we propose a novel scenario in which the photons of the line undergo a radiation transfer process called down-Comptonization after generation by the electron--positron pair annihilation. By incorporating the gamma-ray burst dynamical evolution, we systematically reproduce the observed evolution of the central energy, width, and flux of the emission line and further impose stringent constraints on the production of high-energy emission lines in general. Our study provides a new direction to the research of extreme cosmic bursts.
\end{abstract}
 

\section{Introduction} 
A gamma-ray burst (GRB) is the most violent stellar explosion in the Universe.
 As the brightest-of-all-time \citep{2023ApJ...946L..31B} event triggered at $T_{\rm 0}$ of 2022-10-09T13:17:00 UTC, GRB 221009A has been observed by many telescopes across the full electromagnetic band. Although the extreme brightness of the prompt emission caused trouble in the observation for most gamma-ray instruments, GECAM-C has made accurate and high-resolution measurements of the temporal and spectral properties, thanks to its dedicated designs \citep{2023arXiv230301203A,2024ApJ...972L..25Z}. Together with the redshift $z\,=\,0.151$ \citep{2022GCN.32648....1D,2023arXiv230207891M}, the measured isotropic energy ($E_{\rm iso}$) of GRB 221009A is $1.5\times10^{55}$\,erg \citep{2023arXiv230301203A}, setting a new GRB record.

One of the most surprising and puzzling discoveries is the feature of the MeV emission line shown in the spectrum of GRB 221009A \citep{2024Sci...385..452R,2024SCPMA..6789511Z}. 
 The main properties of the emission line include \citep{2024SCPMA..6789511Z} the following: (1) The central energy of the emission line decays from 37 to 6\,MeV, following a power-law temporal evolution: $E_{\rm c}\propto(t-t_{\rm 0})^{-1.05_{-0.22}^{+0.16}}$, where $t$ is the observed time and $t_{\rm 0}= 226_{-10}^{+8}$\,s. (2) The width-to-central-energy ratio, $\sigma/E_{\rm c}$, is almost constantly about 10\%. (3) The line flux exhibits a temporal decay following a power-law dependence: ${\rm FLux}_{\rm line}\propto (t-t_{\rm 0})^{-2.16_{-0.12}^{+0.09}}$. The discovery of power-law evolution not only reveals critical characteristics of the emission line but also demonstrates that these emission lines must originate from the GRB, rather than from background or instrumental effects \citep{2024SCPMA..6789511Z}.
 
 In order to explain the MeV emission line, some theoretical attempts have been suggested \citep{2024ApJ...973L..51P,2024Sci...385..452R,2024MNRAS.535..982Y,2024ApJ...968L...5W,2024SCPMA..6789511Z,2024ApJ...973L..17Z, 2025ApJ...983L..33Z}. However, none of these models has naturally and completely explained all the observed properties as mentioned above, especially the power-law evolutions of flux and central energy of the emission line. \cite{2024SCPMA..6789511Z} preliminarily proposed that all emission lines at different epochs are originated from the electron--positron ($e^{\pm}$) pair annihilation within dense clumps embedded in the GRB jet, and the dense clumps emitting the MeV line comove with the jet. As a result, the MeV emission line with its evolution is determined by and thus can be used to derive the dynamical properties of the jet. However, this model overestimates the decay rate of the jet Lorentz factor. 
   
 In this work, we present a novel physical scenario that considers both the down-Comptonization effect and the GRB jet dynamics. We systematically resolve the evolution of the central energy, width, and flux of the emission line and further impose stringent constraints on the physical parameters of the down-Comptonization region, with an electron density of $2.0\times 10^9$\,cm$^{-3}$ and an electron temperature of 2.0\,keV. The theoretical understanding, from the aspect of spectral line in the high-energy band, reveals the existence of dense and thermal plasma in GRB, which points out a new direction for the research of the extreme cosmic bursts.
 
\begin{figure*} 
\centering
\includegraphics[width=0.8\textwidth]{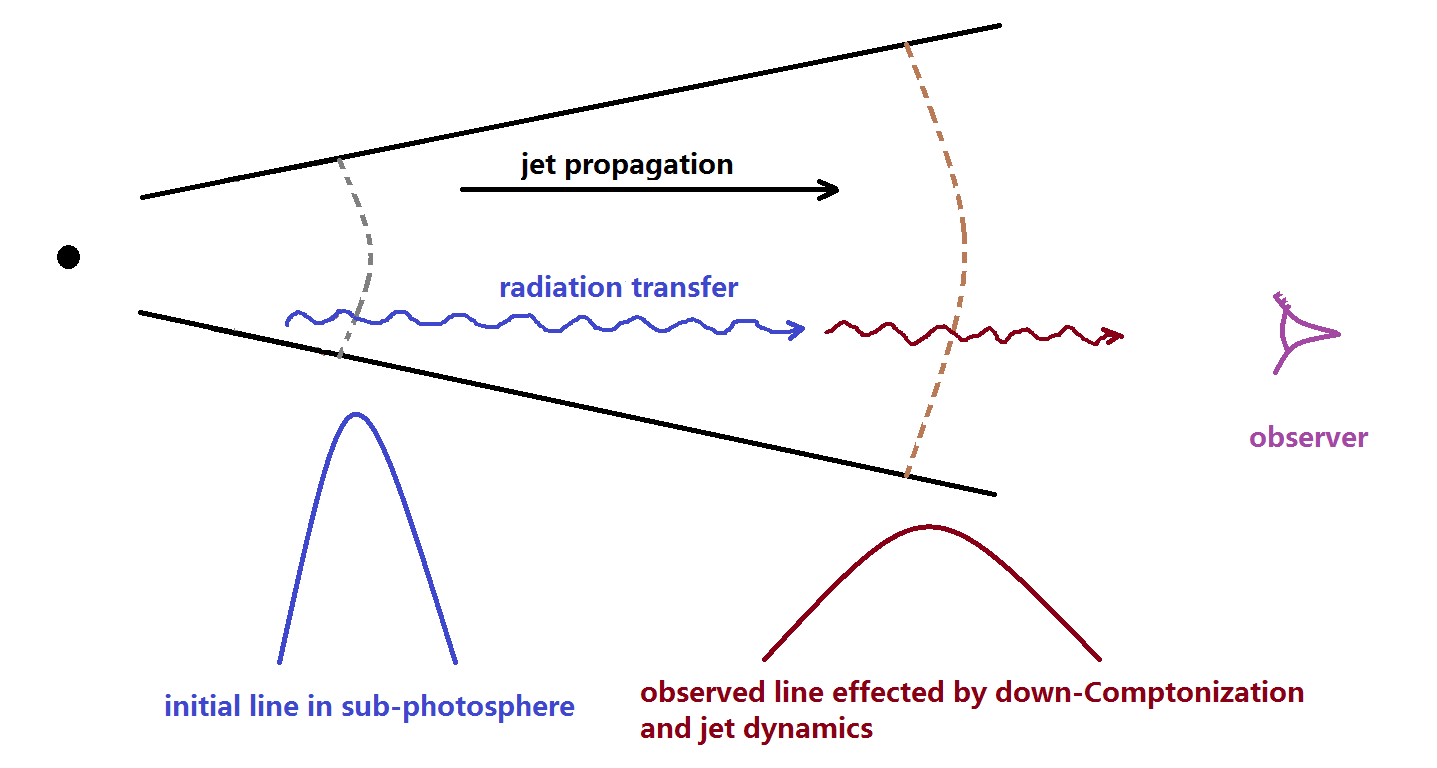}
\caption{The illustration of the MeV emission line released in GRB 221009A. The initial line is originated from electron--positron annihilation in the GRB subphotosphere. During the time of the GRB jet propagation, the emission line interacts with the material inside the jet. This special radiation transfer process is called as the down-Comptonization effect. As time evolves, the emission line is shifted toward the low-energy band and the width of the line is to be broadened. The subphotosphere is drawn in the region labeled by the gray-colored dashed line line, and the down-Comptonization region is shown between the gray-colored dashed line and brown-colored dashed line. }
\label{fig:carton}
\end{figure*}

The remaining part of this work is organized as follows: In Section 2, we present the description of the methods. First, we present the theory of down-Comptonization and conduct calculations regarding the evolution of the emission line under various physical conditions. These conditions encompass variations in temperature and density within the down-Comptonization region.
Afterwards, we carry out an investigation into the dynamical evolution of the jet. In Section 3, we make a comparison between our theoretical results and the observed emission line reported in the work of \cite{2024SCPMA..6789511Z}. We impose constraints on both the dynamical evolution of the jet associated with GRB 221009A and the physical conditions within the down-Comptonization region.
Finally, the conclusions are presented in Section 4.

\section{Methods} \label{sec:methods}
In our scenario, a brief duration of the initial 511\,keV line emission is produced by the electron--positron ($e^{\pm}$) pair annihilation in an inner region of the jet called as subphotosphere \citep{2010MNRAS.407.1033B} of the GRB at redshift of 0.151. Then, the photons of the emission line will naturally undergo interactions with the material in the outer region of the jet during its path to the observer, that is, a special radiation transfer process named down-Comptonization will take place \citep{1978ApJ...219..292R,2003ApJ...593..946K,2004A&A...417..381L,2020ApJ...900...10L,2021A&A...650A..74C}. During the down-Comptonization process, a portion of the emission line's energy is transferred to the material involved in the interaction. This process, in turn, will alter the temporal and spectral characteristics of the line emission observable by the observer. We provide an overall illustration of the process in Figure \ref{fig:carton}. In the following two subsections, we will present the process of down-Comptonization and the jet dynamics in detail.
\subsection{Down-Comptonization}\label{subsec:down-comptonization}

In the GRB research field, the down-Comptonization process was first suggested to apply in the study of the soft X-ray spectrum by \cite{2020ApJ...900...10L}. The results of our previous research indicate that the central energy of the emission line would shift to lower energy after experiencing down-Comptonization with the low-temperature electrons. Furthermore, the line profile deviates from its initial Gaussian shape and eventually reaches thermal equilibrium by displaying a blackbody radiation spectrum. Through numerical calculations, we analyzed the detectability of the X-ray emission line and determined both the temperature and the density of electrons within the down-Comptonization region. 

In this work, we suppose that the original emission line in GRB 221009A will be subject to the down-Comptonization effect when propagating in homogeneous plasma with electron temperature $T_{\rm e}$ and density $N_{\rm e}$ within the jet.
In our model, we assume that the scattering is continuous. Therefore, the spatial transport of radiation can be neglected. The Comptonization process can be well described by the Kompaneets equation. We adopt the extended Kompaneets equation, which reads \citep{2021A&A...650A..74C} 
\begin{eqnarray}\label{e_kompaneets}
\frac{\partial n'}{\partial t'} = \frac{kT_{\rm e}}{m_{e}c^2}N_{\rm e}\sigma_{T}c\frac{1}{x^2}\frac{\partial }{\partial x}\left\lbrace x^4\left(1+\frac{14}{5}\frac{ kT_{\rm e}}{m_{e}c^2}x^2\right)\left(\frac{\partial n'}{\partial x}+n'^2+n'\right)\right\rbrace.
\end{eqnarray}
 
The ratio of $x\equiv\frac{h\nu'}{k T_{\rm e}}$ represents the dimensionless photon energy, where $\nu'$ is the frequency of the photon in the comoving frame, $h=6.63\times10^{-27} {\rm erg\,s} $ is Planck's constant, and $k = 1.38\times10^{-16}$\,erg\,K$^{-1}$ is Boltzmann's constant. The expression $n'(x,t')\equiv n'(\nu',t')$ denotes the photon's occupation number at frequency $\nu'$ and time $t'$ in the comoving frame. It is worth noting that the coefficient of the term of $\frac{kT_{\rm e}}{m_{e}c^2}x^2$ on the right-hand side of the Equation (\ref{e_kompaneets}) is $\frac{14}{5}$, which is different from the value of $\frac{7}{10}$ that in previous works \citep{1994JPhA...27.2905C,2004A&A...417..381L,2020ApJ...900...10L}. \cite{2021A&A...650A..74C} rederived the Kompaneets equation
through expanding the distribution function with respect to the change in the electron’s momentum rather than the electron’s velocity. This newly modified  equation is applicable across a variety of scenarios, including both up-Comptonization (where $h\nu'\gg m_{\rm e}c^2$) and down-Comptonization (where $h\nu'\ll m_{\rm e}c^2$), as well as the intermediate regime, 
where $h\nu'\sim m_{\rm e}c^2$. The energy density of the radiation field is presented as $
u'_{\rm \nu'}= \frac{8\pi\nu'^2}{c^3}n'(\nu',t')h\nu' \propto n'(x,t')x^3$. The spectral intensity can be thus derived as $I'_{\rm \nu'} = \frac{c}{4\pi}u'_{\rm \nu'}\propto n'(x,t')x^3$. Therefore, we can utilize the Equation  (\ref{e_kompaneets}) 
to numerically calculate the evolution of the photon occupation number, as well as the intensity of the emission line.

\cite{2024SCPMA..6789511Z} employed a Gaussian profile to fit the photon number distribution $N(E)$, i.e., $N(E)= A\frac{1}{\sqrt{2\pi}\sigma}{\rm exp} \left[-\frac{(E-E_{\rm line})^2}{2\sigma^2}\right]$, where $A$ is the flux ($\text{photons}\cdot \text{cm}^{-2}~\text{s}^{-1}$) in the line, $E=h\nu$ is the photon energy, $E_{\rm line}$ is the line center energy, and $\sigma$ is the line width in the observer's frame, respectively. Consequently, the intensity of the emission line is $I(E) \propto E\cdot N(E) \propto \frac{E}{\sigma}{\rm exp} \left[-\frac{(E-E_{\rm line})^2}{2\sigma^2}\right]$. Therefore, the normalized intensity in the comoving frame in our current work is assumed to be
\begin{equation}
I'(x,0) = \frac{x}{x_{\rm 0}}{\rm exp} \left[-\frac{(x-x_0)^2}{2\sigma_{\rm 0}'^2}\right].
\end{equation}
The photon occupation number of the original emission line is written as
\begin{equation}\label{e:guass}
n'(x,0)\propto x^{-3}I'(x,0)\sim x^{-2} {\rm exp} \left[-\frac{(x-x_0)^2}{2\sigma_{\rm 0}'^2}\right].
\end{equation}
 The central energy of the emission line is $E_{\rm c}'=h\nu'_0=511\,{\rm keV}$, and the corresponding dimensionless energy is $x_0=h\nu'_0/kT_{\rm e}=511\,{\rm keV}/kT_{\rm e}$. In \cite{2024SCPMA..6789511Z}, it was found that the observed width-to-central-energy ratio remains 10\% from $T_{\rm 0}$+246\,s to $T_{\rm 0}$+320\,s. Therefore, the parameter $\sigma_{\rm 0}'$ is assumed to be $0.1x_{\rm 0}$.

\begin{figure}[ht!]
\centering
\includegraphics[width=0.5\textwidth ]{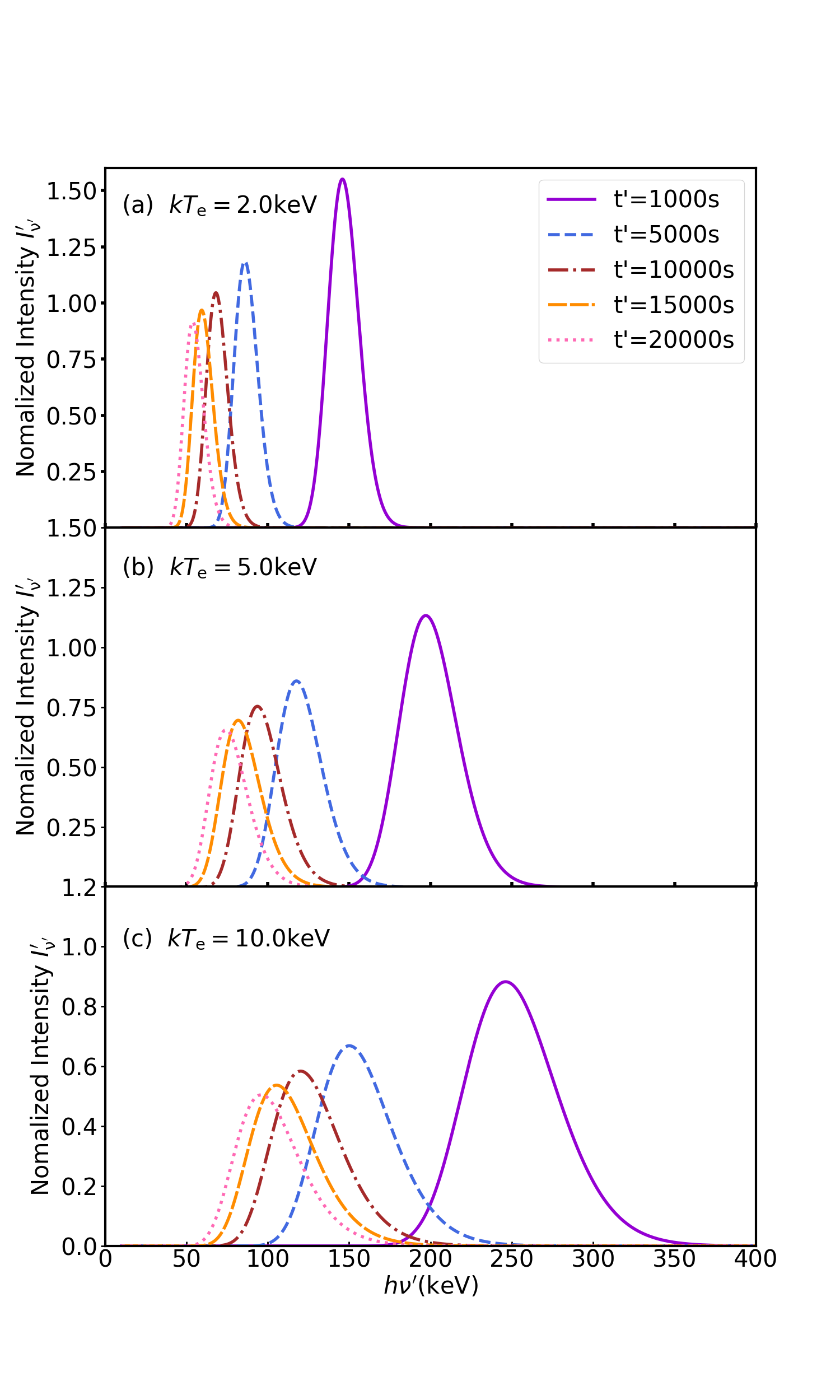} 
\caption{The effect of electron temperature on the evolution of the emission line in the comoving frame. The electron temperatures are illustrated as follows: 2.0\,keV in panel (a); 5.0\,keV in panel (b); and 10.0\,keV in panel (c). The density in all three panels is equivalent to $1.0\times10^9{\rm cm}^{-3}$.}
\label{fig:nu-inu-te}
\end{figure}
   
 We first investigate the impact of the temperature of electron on the temporal evolution of emission lines and fix the fiducial electron density $N_{\rm e}$ to $1.0\times10^9\,{\rm cm}^{-3}$. It is a characteristic density in the jet and less than that in the region of $e^{\pm}$ pair annihilation, which is on the order of $10^{12}$--$10^{13}$\,cm$^{-3}$ \citep{2024Sci...385..452R,2024SCPMA..6789511Z}. We calculate the effects of down-Comptonization on the emission lines using three distinct electron temperatures (2.0, 5.0,  and 10.0\,keV) as examples. In Figure \ref{fig:nu-inu-te}, the evolution of emission line profiles at five specific epochs in the comoving frame at $t'$=1,000\,s; 5,000\,s; 10,000\,s; 15,000\,s; and 20,000\,s is shown. Each line of distinct colored line style  represents the emission line resulting from the interaction between the initial 511\,keV line photons and low-temperature electrons over a specific time. From each panel, it can be seen that the central energy of the emission line decreases noticeably over time as its energy is transferred to electrons. A comparison of the three panels in Figure \ref{fig:nu-inu-te} demonstrates that the central energy of the emission line decreases much more dramatically under lower-temperature conditions at the same time. Furthermore, slightly asymmetrical profiles, with the left wing of the emission line showing a slope relatively steeper than the right wing, can also be observed in Figure \ref{fig:nu-inu-te} during the later stages of evolution, as indicated by the orange dashed and pink dotted lines.

\begin{figure}[ht!]
\centering
\includegraphics[scale=0.3]{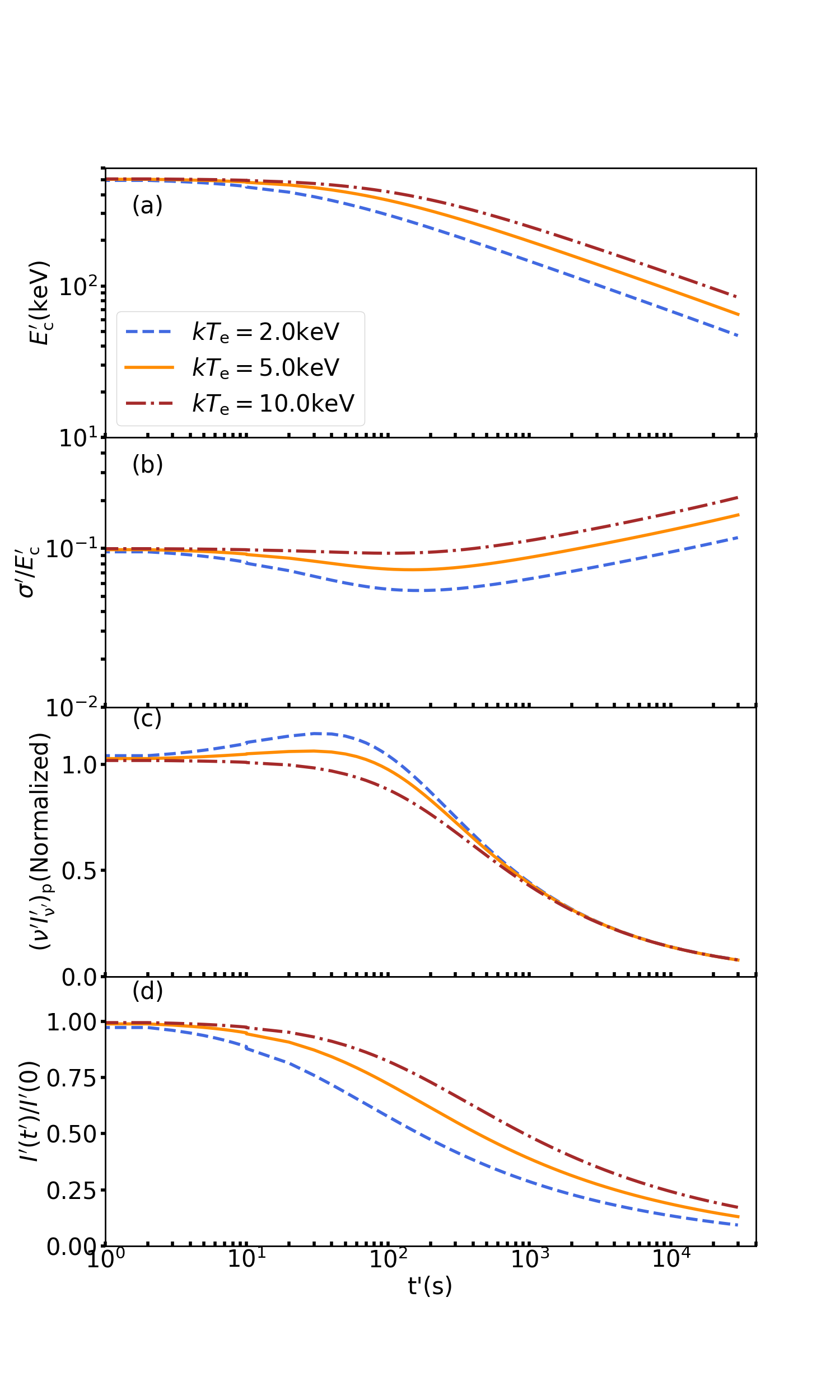} 
\caption{The effect of electron temperature on the evolution of the emission line's central energy $E'_{\rm c}$ (panel (a)), the ratio of $\sigma'/E'_{\rm c}$ (panel (b)), the normalized peak intensity of $(\nu' I'_{\rm \nu'})_{\rm p}$ (panel (c)), and the relative intensity $I'(t')/I'(0)$ (panel (d)). Electron temperatures of 2.0, 5.0, and 10.0\,keV are represented by the blue dashed line, orange solid line, and brown dashed-dotted line, respectively. The density in all three panels is equivalent to $ 1.0\times10^9{\rm cm}^{-3}$.}
\label{fig:sigma-eline-eIe-te}
\end{figure}

We model the photon number distribution $x^2n'(x,t')$ using a Gaussian function of the form $g(x,t')=A_{\rm 0}\cdot{\rm exp}[-(x-x_{\rm c})^2/2\sigma_{\rm t'}^2]$, where the fitting is performed via the \textit{curve\_fit} routine from Python's \textit{scipy.optimize} module. This procedure yields the time-dependent parameters $A_{\rm 0}$, $x_{\rm c}$, and $\sigma_{\rm t'}$ at each time $t'$. Consequently, we derive the temporal evolution of both the central energy $E_{\rm c}'=x_c\cdot kT_{\rm e}$ and the width-to-central-energy ratio $\sigma'/E'_{\rm c} = \sigma_{t'}/x_c$. The effects of temperature on the trends of central energy $E'_{\rm c}$, the width-to-central-energy ratio $\sigma'/E'_{\rm c}$, the normalized peak intensity $(\nu' I_{\rm \nu'}')_{\rm p}$, and the relative intensity $I'(t')/I'(0)$ over time are shown in Figure \ref{fig:sigma-eline-eIe-te}. Different colored line styles indicate different electron temperatures within the down-Comptonization region. In panel (a), we find that the central energy decreases over time. The lower the temperature, the larger the decrease in central energy at the same time. For example, at 1,000\,s (indicated by the violet solid line in Figure \ref{fig:nu-inu-te}), the central energies are 145.92, 197.20, and 246.31\,keV for electron temperatures of 2.0, 5.0, and 10.0\,keV, respectively. In panel (b), the temporal evolution of the width-to-central-energy ratio $\sigma'/E'_{\rm c}$ is illustrated. It can be found that, for the case of $kT_{\rm e}$=10.0\,keV (brown dashed-dotted line),
the ratio $\sigma'/E_{\rm c}'$ remains at its initial value of 0.1 for 1,000\,s. Beyond this time, the ratio progressively increases over time. In contrast, for the cases of $kT_{\rm e}=$ 5.0\,keV and $kT_{\rm e} =$ 2.0\,keV, the width of the emission line initially decreases slightly before increasing over time. Panel (c) depicts the temporal evolution of the normalized peak intensity $(\nu' I'_{\rm \nu'})_{\rm p}$, which can be considered as a scaled peak flux. In general, three lines exhibit a decreasing trend over time. Before 40\,s, $(\nu' I'_{\rm \nu'})_{\rm p}$ changes little in the cases of $kT_{\rm e} = 5.0$\,keV and $kT_{\rm e}= 10.0$\,keV. However, in the case of $kT_{\rm e} = 2.0$\,keV, the value of $(\nu' I'_{\rm \nu'})_{\rm p}$ increases slightly from the beginning to 40\,s and then decreases with time. Since the electron temperature in the down-Comptonization region is lower than that corresponding to the thermal broadening of the initial emission line, in a short period of time, the thermal temperature of the emission line might be strongly decreased, and the emission line looks very narrow as shown in panel (b) of Figure \ref{fig:sigma-eline-eIe-te}. However, the energy of the emission line might not be effectively transferred to electrons in such a short period of time. 
Therefore, the radiation intensity corresponding to the center of the emission line is slightly higher than the initial value, which can also be found in panel (a) of Figure \ref{fig:nu-inu-te}. Despite this specific detail  shown in panel (c) of Figure Figure \ref{fig:sigma-eline-eIe-te}, in general, the intensity of the entire emission line decreases over time, which is presented in panel (d) of Figure \ref{fig:sigma-eline-eIe-te}. In panel (c), after 1000\,s, the evolution trends of $(\nu' I'_{\rm \nu'})_{\rm p}$ are similar, regardless of differences in the electron temperature. Panel (d) depicts the temporal evolution of the relative intensity $I'(t')/I'(0)$. It decreases with time as the energy of photons is transported to electrons during efficient scattering.

 \begin{figure}[b]
 \centering
 \includegraphics[scale=0.3]{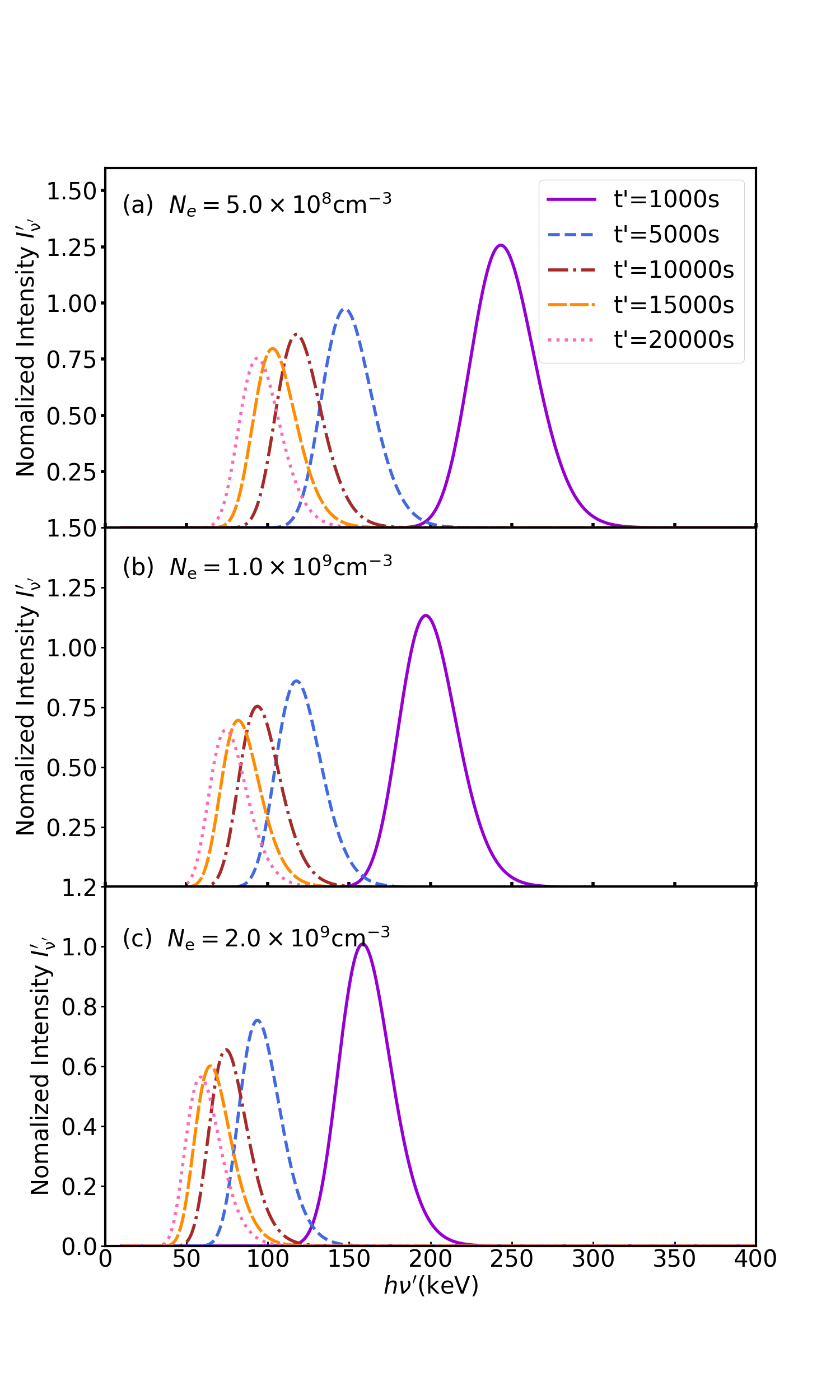} 
 \caption{The effect of electron density on the evolution of the emission line in the comoving frame. Three densities are illustrated as follows: $N_{\rm e}=5.0\times10^{8}{\rm cm^{-3}}$ in panel (a); $N_{\rm e}=1.0\times10^{9}{\rm cm^{-3}}$ in panel (b); and $N_{\rm e}=2.0\times10^{9}{\rm cm^{-3}}$ in panel (c). The electron temperature in all three panels is equivalent to 5.0\,keV.}
\label{fig:nu-inu-ne}
\end{figure}

\begin{figure}[b!]
\centering
\includegraphics[scale=0.3]{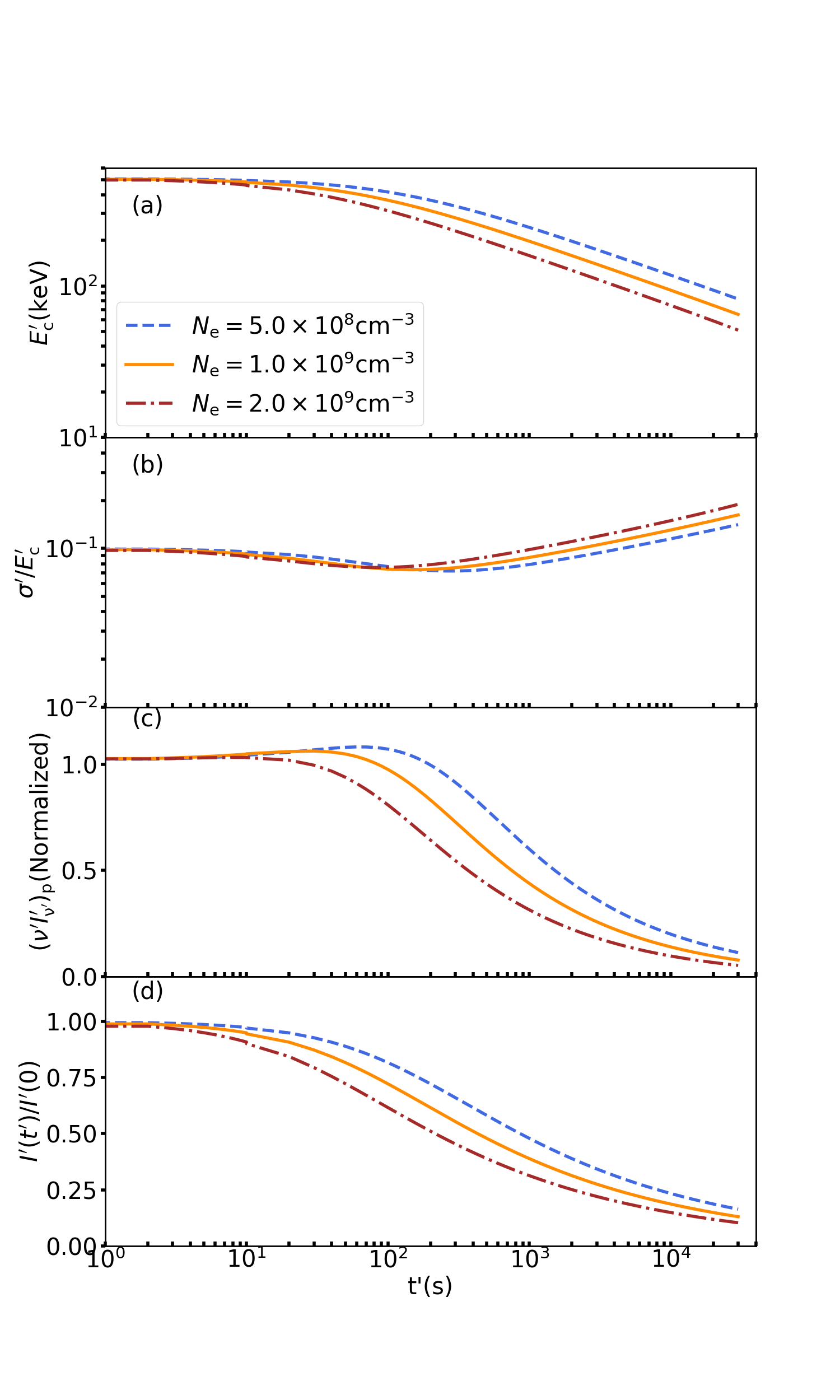} 
\caption{The effect of electron density on the evolution of the emission line's central energy $E'_{\rm c}$ (panel (a)), the ratio of $\sigma'/E'_{\rm c}$ (panel (b)), the normalized peak intensity of $(\nu' I'_{\rm \nu'})_{\rm p}$ (panel (c)), and the relative intensity $I'(t')/I'(0)$ (panel (d)). Three specific densities $N_{\rm e}$ of $5.0\times10^{8}\,\rm{cm^{-3}} $, $1.0\times10^{9}\,\rm{cm^{-3}}$, and $2.0\times10^{9}\,\rm{cm^{-3}}$ are indicated by blue dashed line, orange solid line, and brown dashed-dotted line, respectively. The electron temperature in all three panels is equivalent to 5.0\,keV.}
\label{fig:sigma-eline-eIe-ne}
\end{figure}

The impact of the electron density on the evolution of the emission line is also investigated. The fiducial temperature is fixed at 5.0\,keV. Three cases in the density of $5.0\times10^8\,\rm{cm^{-3}}$, $1.0\times10^9\,\rm{cm^{-3}}$, and $2.0\times10^{9}\,\rm{cm^{-3}}$ are compared, and the results are shown in Figure \ref{fig:nu-inu-ne}. Generally, a higher electron density implies a higher probability of scattering during the same time period. Therefore, the emission energy is efficiently transported more to electrons in higher-density cases.
For example, as shown by the dark violet solid lines in Figure \ref{fig:nu-inu-ne} at 1,000\,s, the central energies are reduced to 243.27\,keV in the case of $N_{\rm e} =5.0\times10^8{\rm cm}^{-3}$, 197.20\,keV in the case of $N_{\rm e} =1.0\times10^9{\rm cm}^{-3}$, and 158.44\,keV in the case of $N_{\rm e} =2.0\times10^{9}{\rm cm}^{-3}$. As time passes, the scattering between electrons and photons becomes increasingly efficient, leading to a gradual shift of the central energy of the emission line toward lower energy after 1000\,s. 

 We also plot the trends of the central energy $E'_{\rm c}$, the width-to-central-energy ratio $\sigma'/E'_{\rm c}$, the normalized peak intensity $(\nu' I'_{\rm \nu'})_{\rm p}$, and the relative intensity $I'(t')/I'(0)$ over time, influenced by the electron density, in Figure \ref{fig:sigma-eline-eIe-ne}. It is evident that variations in density have a significant impact on the evolution of the emission line. In Figure \ref{fig:sigma-eline-eIe-ne}, it is clear that the ratio of $\sigma'/E'_{\rm c}$, the normalized peak intensity $(\nu' I'_{\rm \nu'})_{\rm p}$, and the relative intensity $I'(t')/I'(0)$ are marginally reduced before 10\,s for the three cases. After this time, the figure shows that the central energy $E'_{\rm c}$, the peak intensity $(\nu' I'_{\rm \nu'})_{\rm p}$, and the intensity ratio $I'(t')/I'(0)$ are reduced much more in the case of a higher value of the electron density. However, the width-to-central-energy ratio $\sigma'/E'_{\rm c}$ increases more with larger electron density, as shown in panel (b) of Figure \ref{fig:sigma-eline-eIe-ne}. 

In summary, our theoretical calculations show that the down-Comptonization process has three main effects on the emission line: (1) redshift of the line's central energy; (2) reduction of the line
intensity; and (3) broadening of the line width in the later stage of evolution. It is noted that, without consideration of the redshift effect on the emission line caused by down-Comptonization, arbitrarily comparing the observed central energies of the emission lines with the intrinsic central energy of the line (i.e., 511\,keV) will lead to an overestimation of the evolution behavior of the jet Lorentz factor.

\subsection{Jet Dynamics} 
The MeV emission line moving alone with the GRB jet suffers the down-Comptonization effect. In order to completely explain the time evolution of the MeV emission line in GRB 221009A, we incorporate the dynamical evolution of the jet to the down-Comptonization process. 

The GRB hydrodynamic evolution has been well studied. For example, a comprehensive investigation was proposed by \cite{1999ApJ...515L..49D}. In this framework, we can perform the detailed calculation on the temporal evolution of the bulk motion of the jet \citep{2001ChJAA...1..349M,2001ChJAA...1..433M}. The density profile of the medium surrounding the jet is assumed to be $n(r) = n_{\rm 0}(r/r_{\rm 0 })^{-s}$,
 where $r_{\text 0}$ is the initial radius of the fireball becoming optically thin for producing gamma rays. The value of the power-law index $s$ for the homogeneous case is set to $0$, and in the wind case, it is equal to $2.0$ \citep{1998MNRAS.298...87D}. The dynamical evolution of the Lorentz factor along the radius can be determined as \citep{2001ChJAA...1..349M} 

\begin{eqnarray}\label{e:dgammadchi} 
\frac{d\Gamma}{d\chi}=-(3+g-s)\frac{\chi^{g+2-s}}{\Gamma_0}(\Gamma^2-1)^{\frac32}(\Gamma+1)^{-\varepsilon_{\rm e}}(\Gamma_0^2-1)^{-\frac12}(\Gamma_0+1)^{\varepsilon_{\rm e}},
\end{eqnarray}
where $\chi = r/R_{\rm 0}$ is the radius scaled, $\Gamma_{\rm 0}$ is the initial Lorentz factor, $R_{\rm 0} = \left[(3+g-s)M_{\rm 0}/(\Gamma_{\rm 0}\Omega_{\rm 0}n_{\rm 0}m_{\rm p}r_{\rm 0}^{s-g})\right]^{\frac{1}{3+g-s}}$ is the deceleration radius, and $M_{\rm 0} =4\pi r_{\rm 0}^2 n_{\rm 0} m_{\rm p}/\sigma_{T}$ is the initial mass of the surrounding material. The parameter $\varepsilon_{\rm e}$ represents the fraction of kinetic energy injected into nonthermal electrons. In the case of radiative limit, the value of $\varepsilon_{\rm e}$ is equal to 1. For the adiabatic
limit, $\varepsilon_{\rm e}$ is equal to 0. The initial solid angle of the jet is $\Omega_{\rm 0}= 2\pi(1- \text{cos}\,\theta_{\rm 0})$, where $\theta_{\rm 0}$ refers to the initial half-opening angle. The parameter $g$ represents the degree of the jet beaming effect, determining the solid angle in radius $r$ as $\Omega_{j} =(r/r_{\rm 0})^g\Omega_{\rm 0} $. The measurement of the jet's Lorentz factor is usually influenced by the viewing angle between the direction of the jet and the line sight of the observer. However, in the case of GRB 221009A, the jet's opening angle ($\theta_{\rm j}$) is extremely small ($\theta_{\rm j} \sim 1^\circ$), as reported by recent studies \citep{2023arXiv230301203A,2023Sci...380.1390L,2024ApJ...973L..17Z}. 
Since the radiation is highly collimated within this narrow opening angle, here we simply assume that the viewing angle between the jet's direction and the observer'\,s line of sight is zero.

Combining Equation (\ref{e:dgammadchi}) with the relationship between radius $r$ and arrival time $t$ in \cite{1997ApJ...489L..37S}, namely, $r = 8\Gamma^2ct$ and $dt = dr / 2\Gamma^2c$, we can determine the evolution of $\Gamma$ over time. The value of $r_{\rm 0}$ of GRB 221009A is set to $1.0\times 10^{16}$\,cm \citep{2024ApJ...973L..17Z}. In this paper, we adopt the initial Lorentz factor $\Gamma_{\rm 0} = 560$, the number density in the circumburst medium $n_{\rm 0}=0.4\,{\rm cm^{-3}}$, and the half-opening angle $\theta_{\rm } = 0.^{\circ}6$ as our fiducial parameters. The fiducial values of the power-law index $s$, the beaming factor $g$, and the energy fraction parameter $\epsilon_{\rm e}$ are set as 0.0, 2.0, and 0.0, respectively. The parameter setting for jet dynamics is consistent with the modeling of the multiwavelength observations of GRB 221009A \citep{2023Sci...380.1390L,2024ApJ...973L..51P,2024ApJ...962..115R,2024ApJ...966..141Z}.

\begin{figure*}[b]
\centering
\includegraphics[scale=0.5 ]{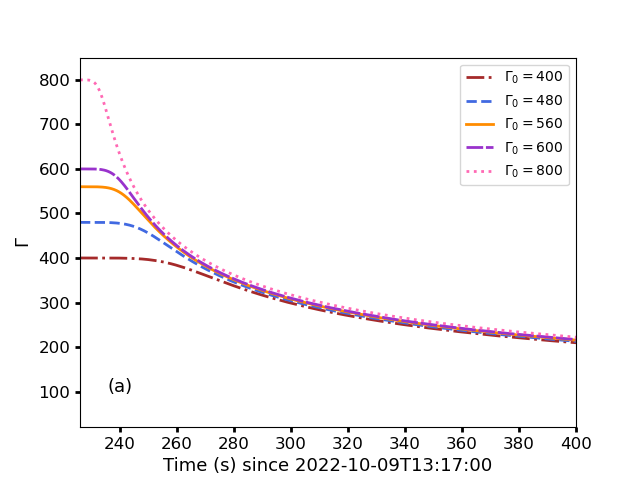}
\includegraphics[scale=0.5 ]{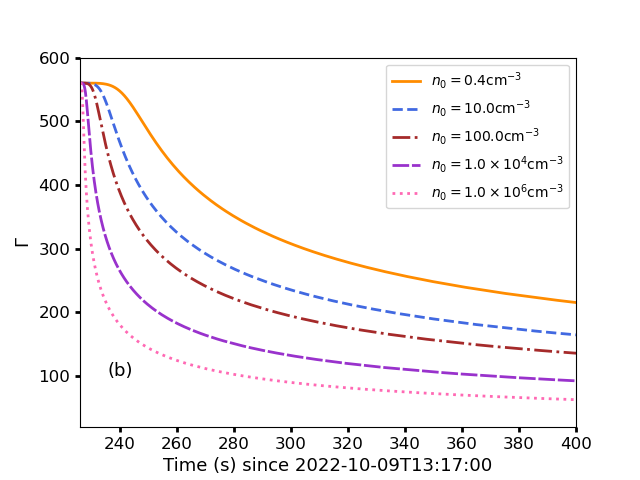}%
\includegraphics[scale=0.5]{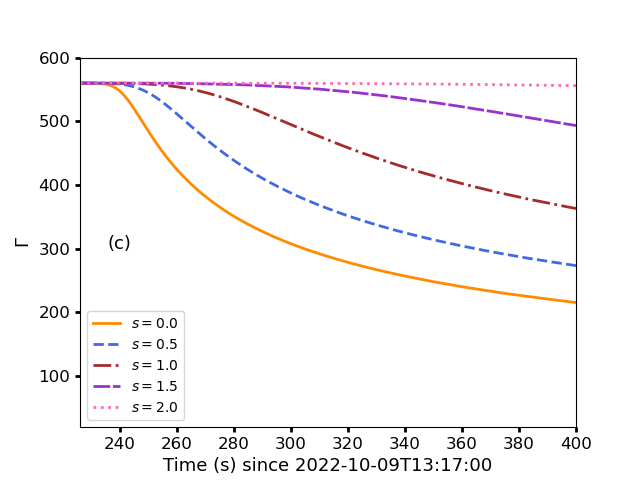}
\includegraphics[scale=0.5 ]{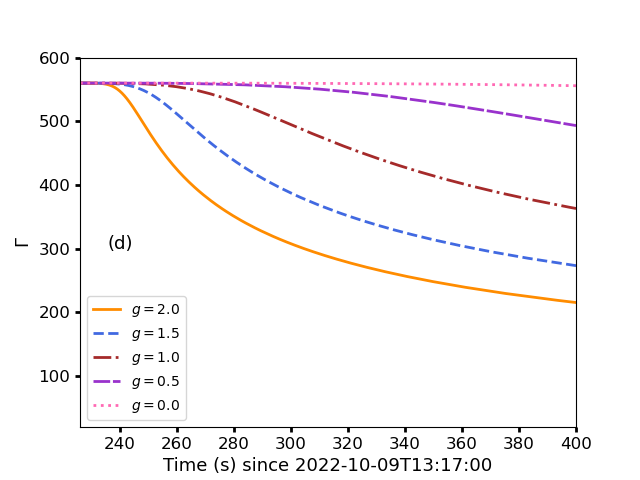}%
\includegraphics[scale=0.5]{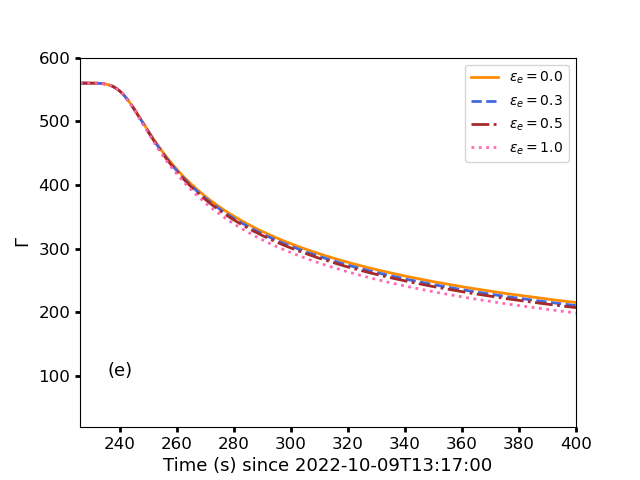}
\caption{Effects of various parameters of jet dynamics on Lorentz factor evolution. The temporal evolution of the Lorentz factor is influenced by these parameters: 
initial Lorentz factor $\Gamma_{\rm 0}$ (panel (a)), density $n_{\rm 0}$ (panel (b)), power index $s$ (panel (c)), beaming factor $g$ (panel (d)), and energy fraction $\epsilon_{\rm e}$ (panel (e)).} \label{fig:gamma-t-parameter}
\end{figure*}

 We examine the effect of modulating these parameters on the temporal evolution of the Lorentz factor from $T_{\rm 0}$+226\,s (corresponding to the start time $t \,=\,0$) to $T_{\rm 0}$+400\,s, as shown in Figure \ref{fig:gamma-t-parameter}. In panel (a), five distinct lines represent various initial Lorentz factors. For all five lines, it can be observed that there exist minimal change phases in which the Lorentz factors almost maintain their initial values. Correspondingly, smaller initial Lorentz factors are associated with longer minimal change durations. For the scenario depicted by the brown dashed-dotted line, where the initial Lorentz factor $\Gamma_{\rm 0} = 400$, the minimal change phase persists for 34\,s until $T_{\rm 0}+$260\,s. When $\Gamma_{\rm 0} = 560$, as indicated by the orange line, the Lorentz factor remains above 550 for only 10\,s. In contrast, when $\Gamma_{\rm 0} = 800$, this duration is dramatically reduced to less than 4\,s until $T_{\rm 0}+$230\,s. However, after the minimal change phase, the evolution of the Lorentz factor for the five cases does not show a significant divergence. Specifically, after $T_{\rm 0}+$280\,s, the profiles of these five lines are amenable to a uniform power-law fit, expressed as $\Gamma(t) \propto t^{-0.42}$. This closely approximates the analytical solution for the jet dynamics in the adiabatic case, namely $\Gamma(t)\propto t^{-3 / 8}$ \citep{1998ApJ...497L..17S,2004RvMP...76.1143P}. In panel (b), the temporal evolution of the Lorentz factor $\Gamma$ is illustrated under the influence of five different densities of the surrounding medium. It is natural to find that the decreased rate of the Lorentz factor is less pronounced in environments with lower circumbient densities. The Lorentz factor in the case of $n_{\rm 0} = 1.0\times10^6\,{\rm cm}^{-3}$ exhibits a power-law distribution from the beginning. The effects of varying the power-law index $s$ and the beaming parameter $g$ on the evolution of the Lorentz factor are illustrated in panels (c) and (d), respectively. An increase in $g$ and a decrease in $s$ result in a more prolonged rate of change for the Lorentz factor. In the scenario of $g\,=\,0$ or $s\,=\,2$, it is evident that the Lorentz factor remains unchanged. Panel (e) shows that the energy fraction $\epsilon_{\rm e}$ has a negligible impact on the temporal evolution of the Lorentz factor from $T_{\rm 0}$+226\,s to $T_{\rm 0}$+400\,s.

 From Figure \ref{fig:gamma-t-parameter}, we find that not all the evolution curves of the Lorentz factor follow a power-law distribution from the very beginning of $T_{\rm 0}+226$\,s. Some of them go through a minimal change or a gradual decline phase before conforming to a power-law shape. We utilize the power-law form $B\cdot(t/t_{\rm 10})^{\alpha}$ to model the evolution of the Lorentz factor beyond $t_{\rm 10}$, where $t_{\rm 10}$ is the time at which the Lorentz factor drops to less than 90\% of its initial value. A higher value of $t_{\rm 10}$ implies a longer phase of minimal change or gradual decline.  The fitting values of power-law index $\alpha$ and the mean squared error (MSE) are also obtained using the function of \textit{curve\_fit} in Python. The value of MSE is calculated as the average of the squared differences between the fitted values and the actual values of Lorentz factor derived by dynamical calculation for all time series data points under specific jet'\,s dynamical parameters. The smaller the MSE value, the better the power-law fitting result. As illustrated in panel (e) of Figure \ref{fig:gamma-t-parameter}, the effect of $\epsilon_{\rm e}$ is negligible. Furthermore, increasing the power index $s$ and decreasing the value of $g$ produce a similar evolutionary trend in the Lorentz factor. Consequently, we focus on the distributions of $t_{\rm 10}$ and $\alpha$ in the planes defined by $g$--${\rm log}n_{\rm 0}$, $\Gamma_{\rm 0}$--$g$, and $\Gamma_{\rm 0}$--${\rm log}n_{\rm 0}$. The ranges considered for $g$, ${\rm log}n_{\rm 0}$ and $\Gamma_{\rm 0}$ are [0.0, 2.0], [-2.0, 6.0], and [400.0, 800.0], respectively.

\begin{figure*}[b]
\centering
\includegraphics[scale=0.27]{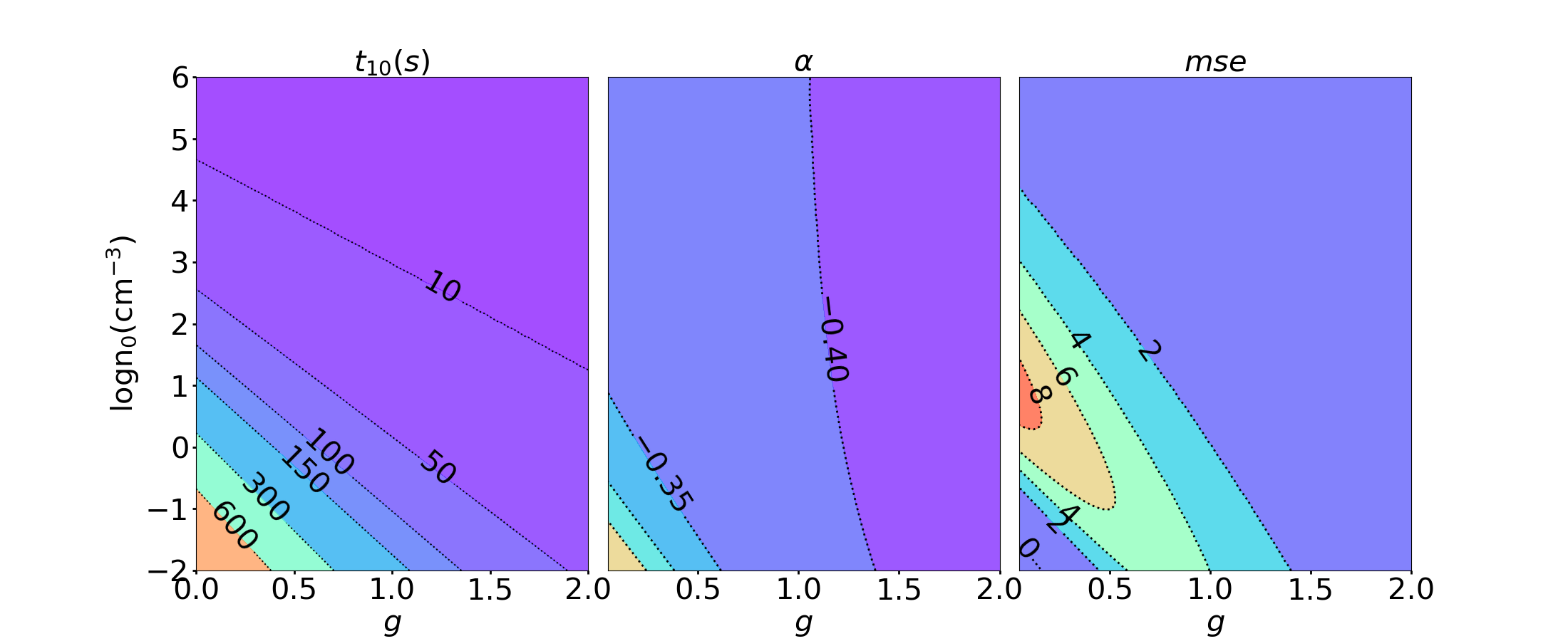}
\includegraphics[scale=0.27 ]{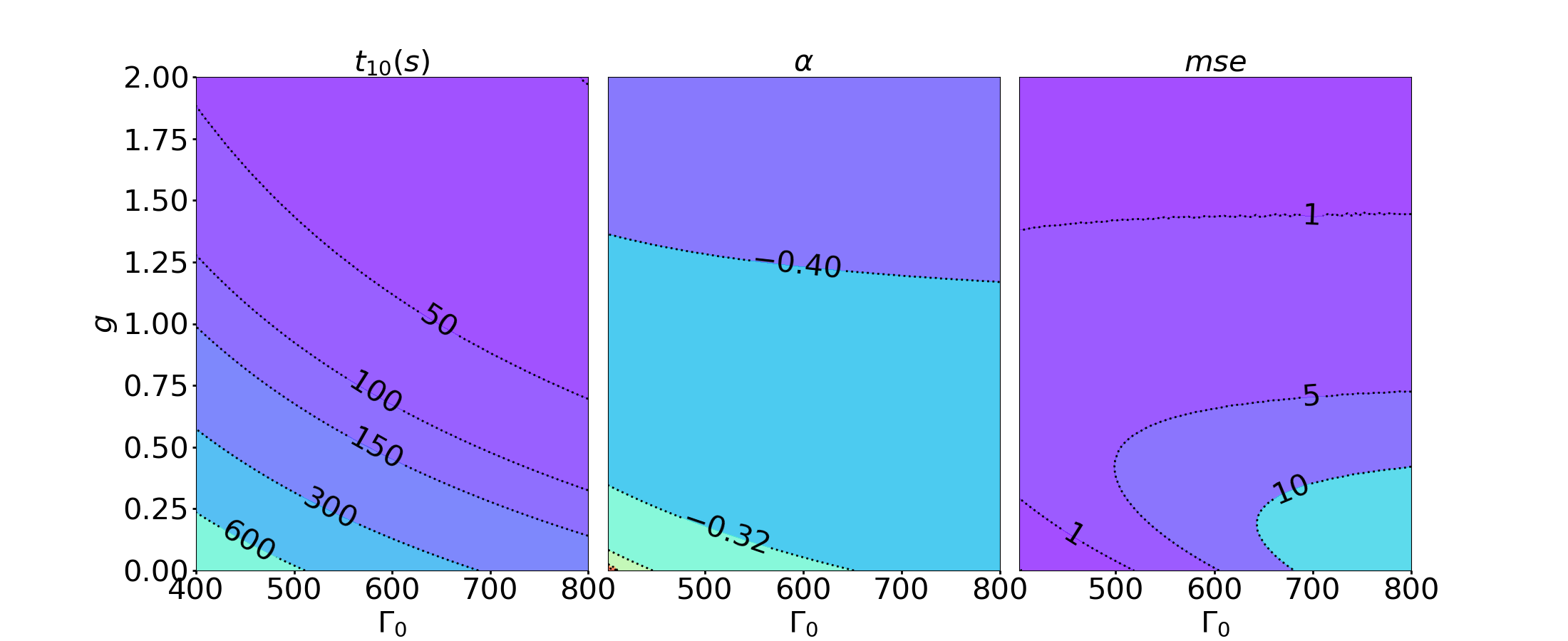}
\includegraphics[scale=0.27]{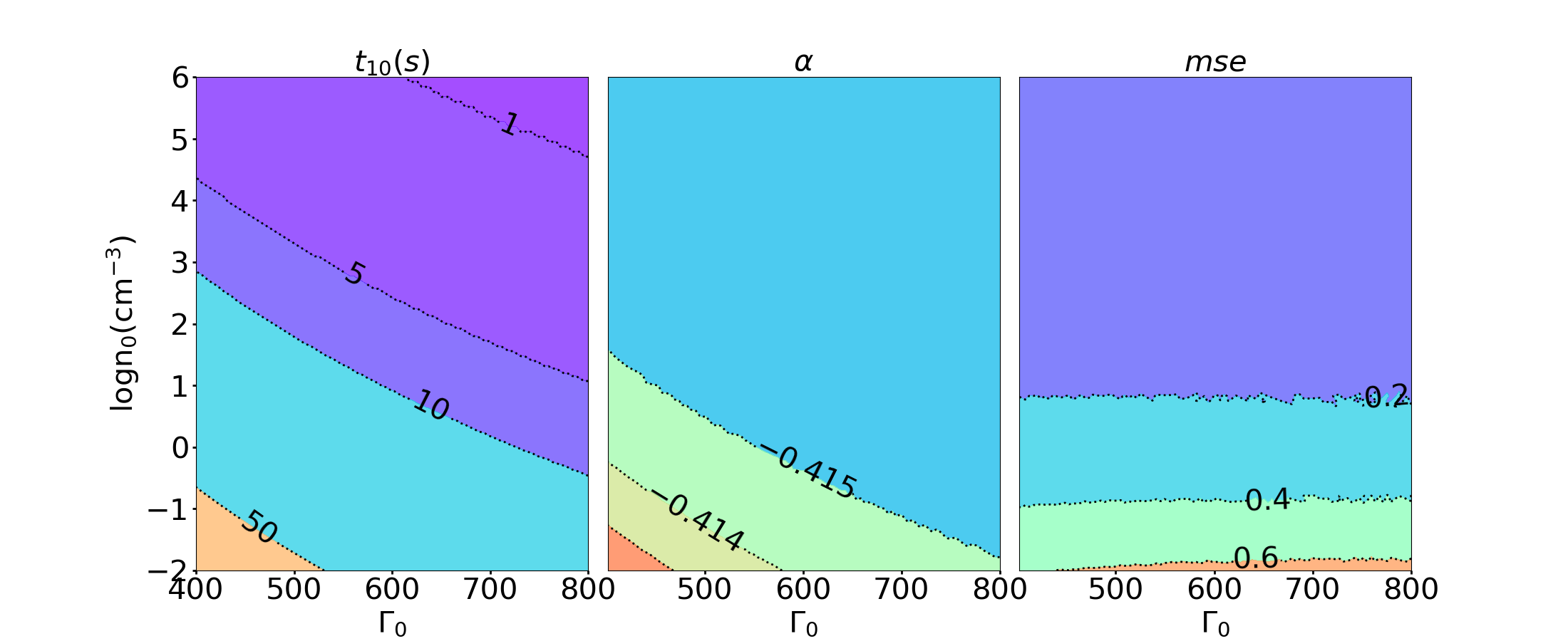}
\caption{The contour distributions of $t_{\rm 10}$, the fitting power-law index $\alpha$, and the MSEs within the parameter planes.}
\label{fig:alpha-t-mse-parameter}
\end{figure*}
 
The contour distributions of the values of $t_{\rm 10}$ and $\alpha$ within the parameters space are presented in the left and central panels of Figure \ref{fig:alpha-t-mse-parameter}, while the corresponding MSE resulting from the fitting of the power law are illustrated in the right panels. Within the $g$-${\rm log}n_{\rm 0}$ plane (top three panels), $t_{\rm 10}$ is found to exhibit a decreasing trend as $g$ and $n_{\rm 0}$ increase. This is also consistent with the evolutionary trends of the Lorentz factor shown in panels (b) and (d) of Figure \ref{fig:gamma-t-parameter}, which demonstrates that the phases of minimal change extend for shorter durations in the cases with higher values of $n_{\rm 0}$ and $g$. A similar trend is also observed within the $\Gamma_{\rm 0}$-$g$ plane, indicating that $t_{\rm 10}$ decreases with increasing $\Gamma_{\rm 0}$. In the bottom-left panel, specifically within the $\Gamma_{\rm 0}$- ${\rm log}n_{\rm 0}$ plane, we find that $t_{\rm 10}$ is significantly reduced, implying that the phases of minimal change are shortened. By integrating the panels in the middle and on the right, it is evident that in regions characterized by lower values of $mse$, the fitted power-law exponents consistently converge around -0.4, confirmed to the results shown in Figure \ref{fig:gamma-t-parameter}.    
 
\section{Results: Observational Constraints} 
The parameters of Gaussian fitting for the line components at each observational epoch, regarding the central energies of emission lines and their corresponding errors, the widths, and the peak fluxes, are listed in columns (2)--(4) of Table \ref{table1}, were obtained from \cite{2024SCPMA..6789511Z}. The eight emission lines at the epochs $T_{\rm 0}$+(246, 256), $T_{\rm 0}$+(270,275), $T_{\rm 0}$+(275, 280), $T_{\rm 0}$+(280, 285), $T_{\rm 0}$+(285, 290), $T_{\rm 0}$+(290, 295), $T_{\rm 0}$+(295, 300), and $T_{\rm 0}$+(300, 320)\,s, are adopted. The time evolution of the Lorentz factor shown by the orange  solid line in Figure \ref{fig:gamma-t-parameter} is considered. This evolutionary behavior of the Lorentz factor corresponds to a homogeneous density of the circumburst material of $0.4\,{\rm cm}^{-3}$ and $s\,=\,0$, an initial Lorentz factor of $\Gamma_{\rm 0} = 560$, the adiabatic case with $\epsilon_{\rm e}=0$, and the jet beaming factor $g\,=\,2$.

We deduce the Lorentz factor at the middle time $t_{\rm mid}$ of each observed epoch and the corresponding time $t' = \int_{T_{\rm 0}+246}^{t_{\rm mid}} \Gamma(t)/(1+z) dt$ for detection emission lines in the comoving frame. The eight corresponding times $t'$ for the middle time of the epochs are calculated as 2151, 9987, 11,574, 13,100, 14,570, 15,992, 17,369, 20,648\,s since the emergence of the initial emission line, respectively. Subsequently, we transform the observed emission line features including $E_{c}$ and $\sigma/E_{\rm c}$ into those, i.e., $E'_{c}$ and $\sigma'/E'_{\rm c}$ in the comoving frame. Considering the errors in the fitting data for the observed MeV emission lines \citep{2024SCPMA..6789511Z}, to constrain the physical conditions of the region of down-Comptonization, we calculate the evolutionary trends of $E'_{c}$ and $\sigma'/E'_{\rm c}$ under the following parameters: electron densities of $N_{\rm e} =1.0\times10^9$, $2.0\times10^9$, $3.0\times10^9$, $4.0\times10^9$, and $5.0\times10^9 {\rm cm}^{-3}$, combined with electron temperatures of 1.0, 2.0, 3.0, 4.0, and 5.0\,keV. The comparison between the transformed observational data in the comoving frame and the theoretical evolutionary trends under these 25 physical conditions is shown in Figure \ref{fig:ec-sigma-comoving-compare}. Each panel examines the influence of varying electron densities ($N_{\rm e}$ ranging from $1.0 \times 10^9$ to $5.0 \times 10^9\,\text{cm}^{-3}$,at  increments   of $1.0 \times 10^9\,\text{cm}^{-3}$) at a constant temperature. Compared to the observational data (indicated by red crosses), it is evident that for each temperature depicted in the left panels of Figure \ref{fig:ec-sigma-comoving-compare}, a specific electron density can be identified that well reproduces the evolution of $E'_{\rm c}$. Subsequently, we verify whether the corresponding evolution of $\sigma'/E'_{\rm c}$ under these temperature and density conditions (as shown in the right panels) aligns with the observed trends. Our analysis reveals that the theoretical predictions for the evolutionary trends of both $E'_{\rm c}$ and $\sigma'/E'_{\rm c}$ align well with the observational data only when the electron density is $N_{\rm e} = 2.0 \times 10^9~\text{cm}^{-3}$ and the electron temperature is $kT_{\rm e} =2~\text{keV}$, as illustrated by the orange solid lines in panels (c) and (d) of Figure \ref{fig:ec-sigma-comoving-compare}. 

\begin{figure*}[htbp]
\centering  
\includegraphics[scale=0.45]{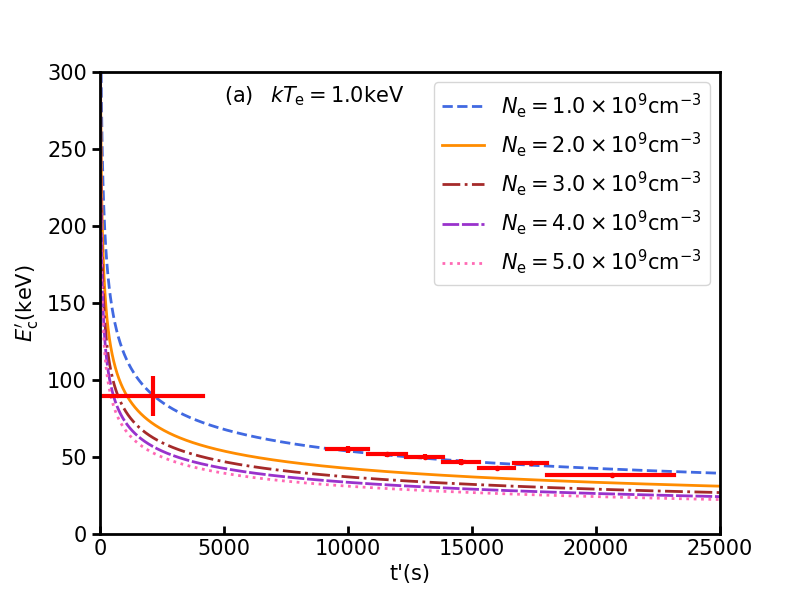}%
\includegraphics[scale=0.45]{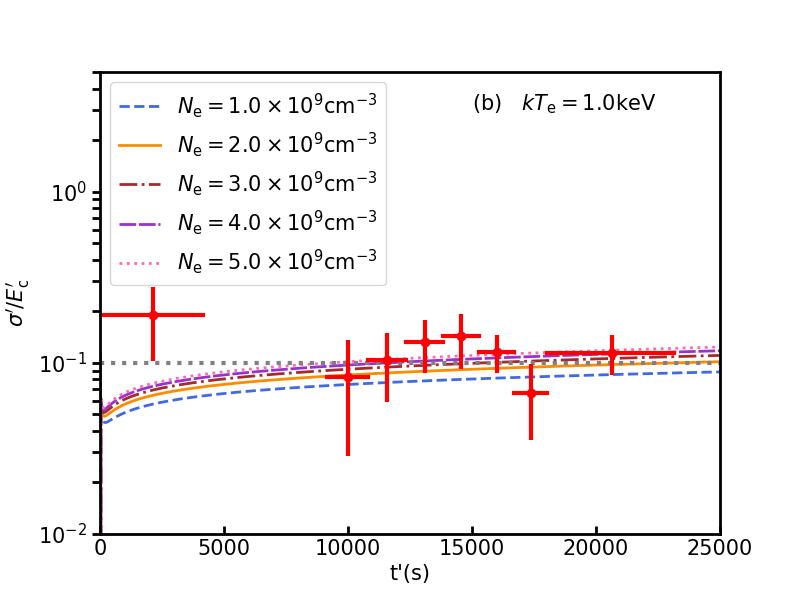}
\includegraphics[scale=0.45]{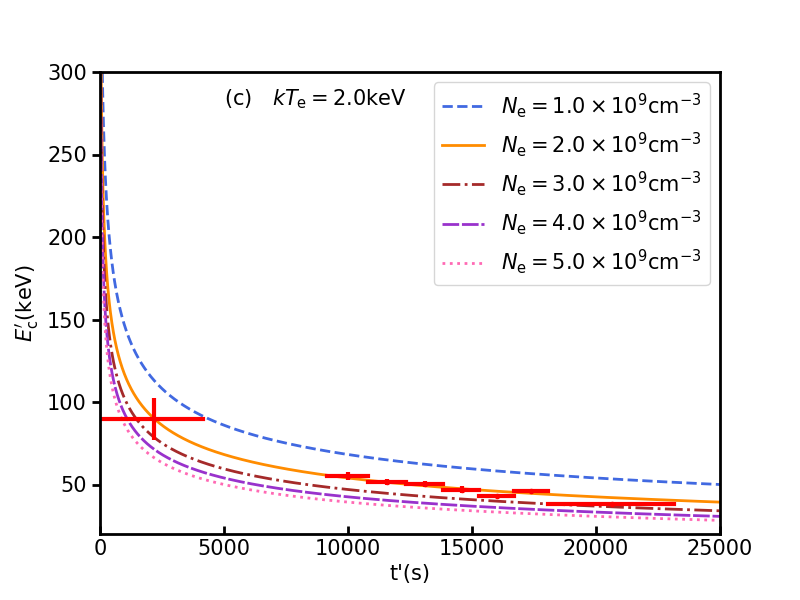}%
\includegraphics[scale=0.45]{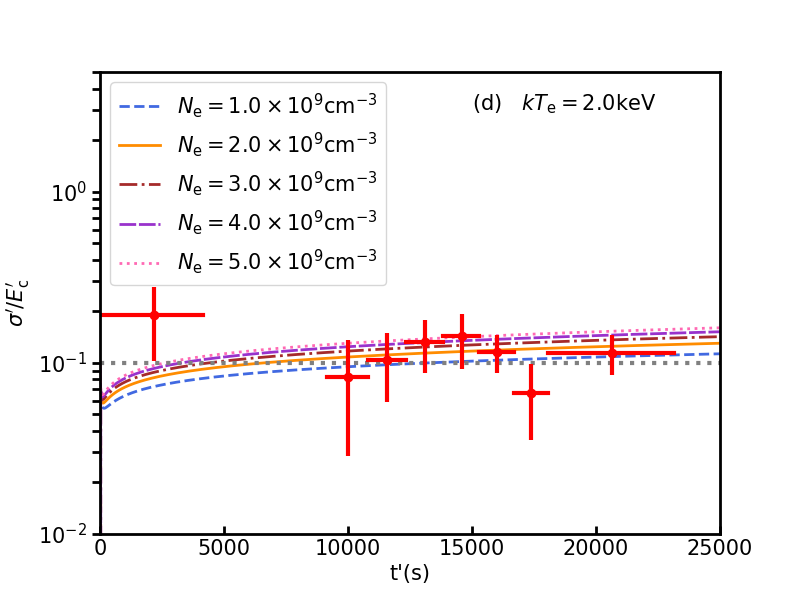}
\includegraphics[scale=0.45 ]{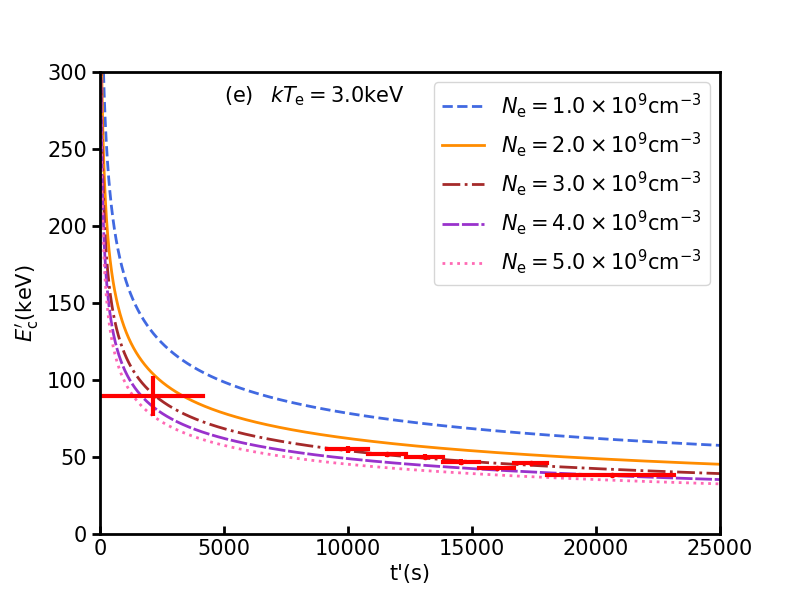}%
\includegraphics[scale=0.45 ]{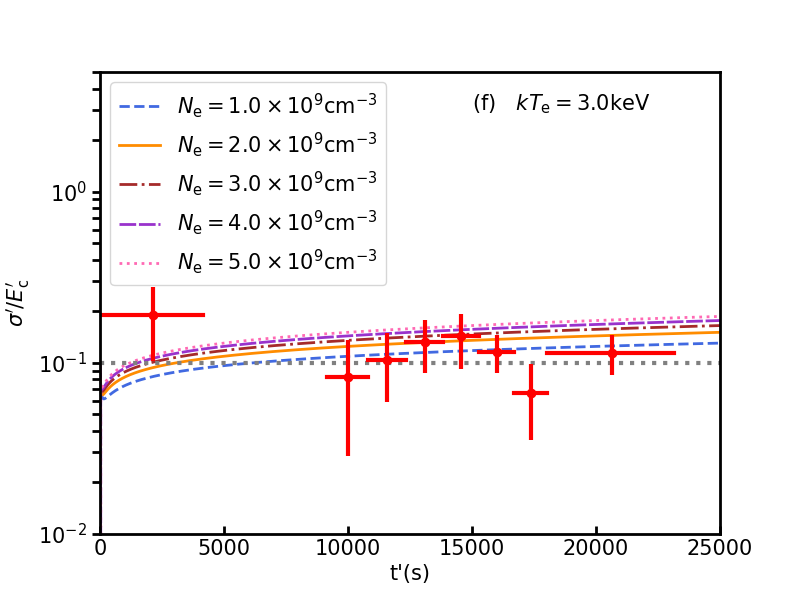}
\caption{The theoretical time evolution of $E'_{\rm c}$ and $\sigma'/E'_{\rm c}$ in the comoving frame within the down-Comptonization region. The horizontal gray dotted lines in the right panels indicate the cases of $\sigma'/E'_{\rm c}=0.1$. The red crosses demonstrate the transferred the observed results in the comoving frame. Within each panel, the effects of five different electron densities (as annotated in the legend) on the evolutionary behavior are compared, while the temperature is held constant.}
\label{fig:ec-sigma-comoving-compare}
\end{figure*}

\begin{figure*}[h]
\centering  
\ContinuedFloat
\includegraphics[scale=0.45]{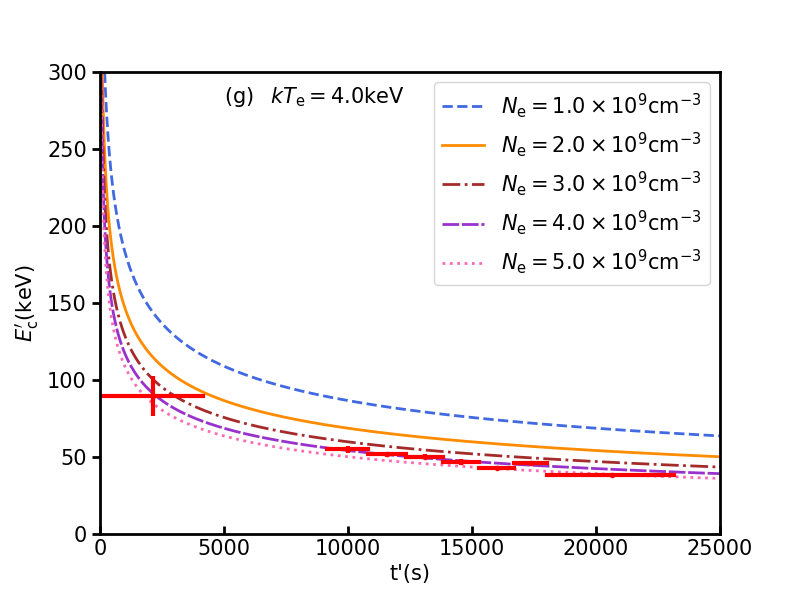}%
\includegraphics[scale=0.45]{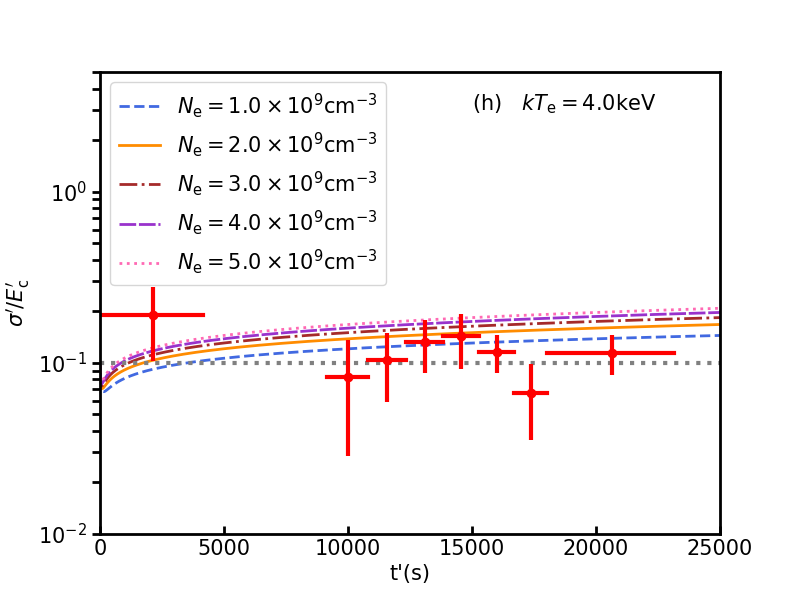}
\includegraphics[scale=0.45 ]{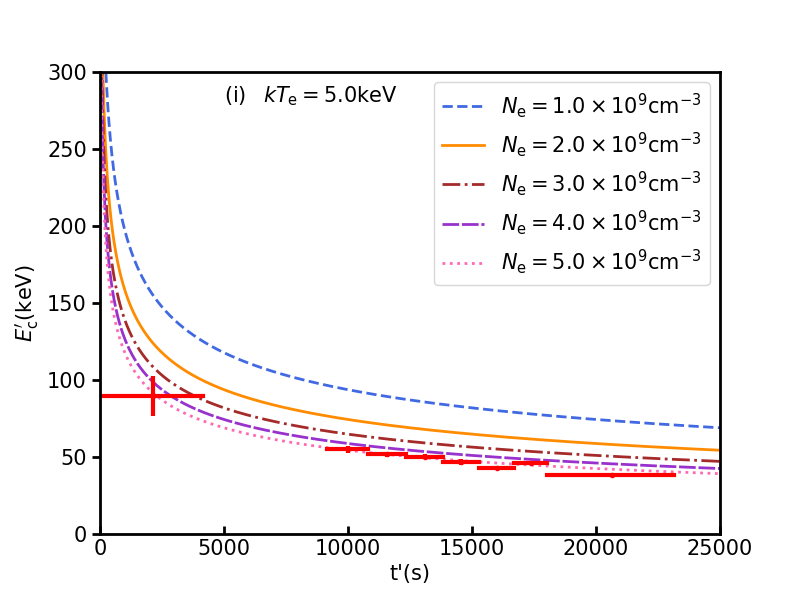}%
\includegraphics[scale=0.45 ]{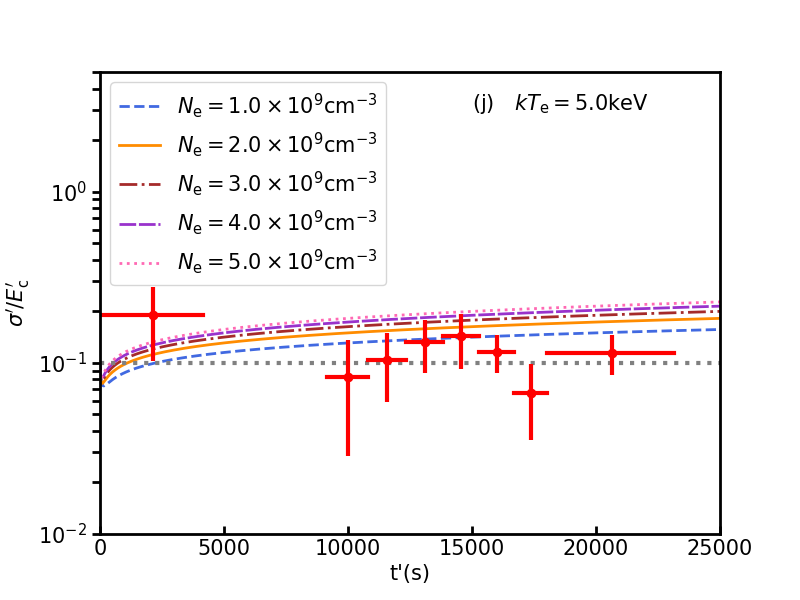}
\caption{(continued) }
\label{fig:continued}
\end{figure*} 

\begin{figure}[htbp]
\centering
\includegraphics[width=0.45\textwidth]{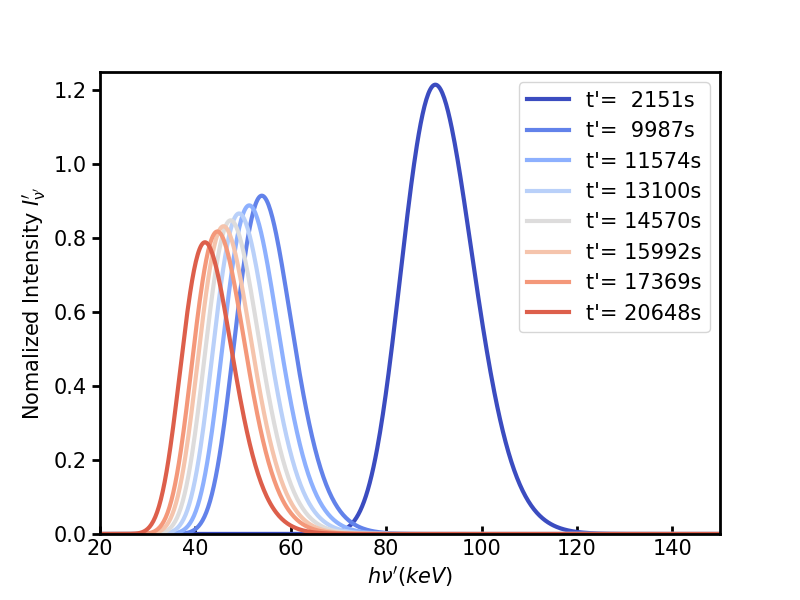} %
\includegraphics[width=0.45\textwidth]{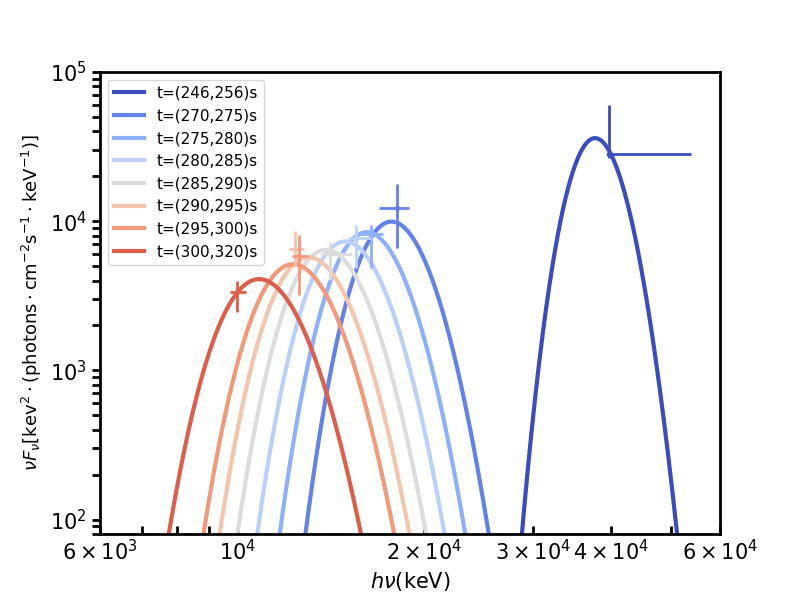}
 \caption{The evolution behaviors of the emission line. The left panel shows the temporal evolution of the normalized intensity $I'_{\rm \nu'}$ in the comoving frame. The right panel shows the theoretical evolution behaviors of fluxes of the emission line in $\nu F_{\rm \nu}$ representation and the observed peak flux in the observer's frame. The emission lines are represented by eight different colors, and each corresponds to a specific observed epoch. The colored  
 crosses represent the observed peak flux, along with its corresponding photon energy within 1$\sigma$ 
 error, for each epoch.}
 
\label{fig:compare1}
\end{figure}

Therefore, we present the intrinsic emission line profiles influenced by the down-Comptonization effect, under the condition of $N_{\rm e} = 2.0 \times 10^9~\text{cm}^{-3}$ and $kT_{\rm e} =2~\text{keV}$, at the eight epochs in the comoving frame, as depicted in the left panel of Figure \ref{fig:compare1}. In this plane, the time $t'$ in the legend indicates the middle time for each epoch after the emergency of the first $e^\pm$ annihilation line at $T_{\rm 0}+246$\,s. It can be seen that the central energy of the emission line is significantly redshifted to less than $90.0$\,keV due to the down-Comptonization effect.
According to $I_{\rm \nu} = \delta^3(t) I'_{\rm \nu'}$, where $\delta(t) = \Gamma(t)\,/(1+z)$, the intensity $I'_{\rm \nu'}$ in the comoving frame is transformed into $I_{\rm \nu}$ in the observer's frame, and the observed flux density $F_{\rm \nu}$ at frequency $\nu$ is obtained. Comparison of emission line profiles with observed data is shown in the right panel of Figure \ref{fig:compare1}. Since our model accurately reproduces both the central energy $E_{\rm c}$ and the width $\sigma$ of the observed emission line, we display only the peak spectral energy flux $(\nu F_{\rm \nu})_{\rm p}$ in Figure \ref{fig:compare1} for clarity, omitting the full line profiles. The corresponding observed peak flux $(\nu F_{\rm \nu})_{\rm p}$, along with its 1$\sigma$ error marked by colored crosses, is also shown.

\begin{figure}[htbp]
\centering
\includegraphics[scale=0.25]{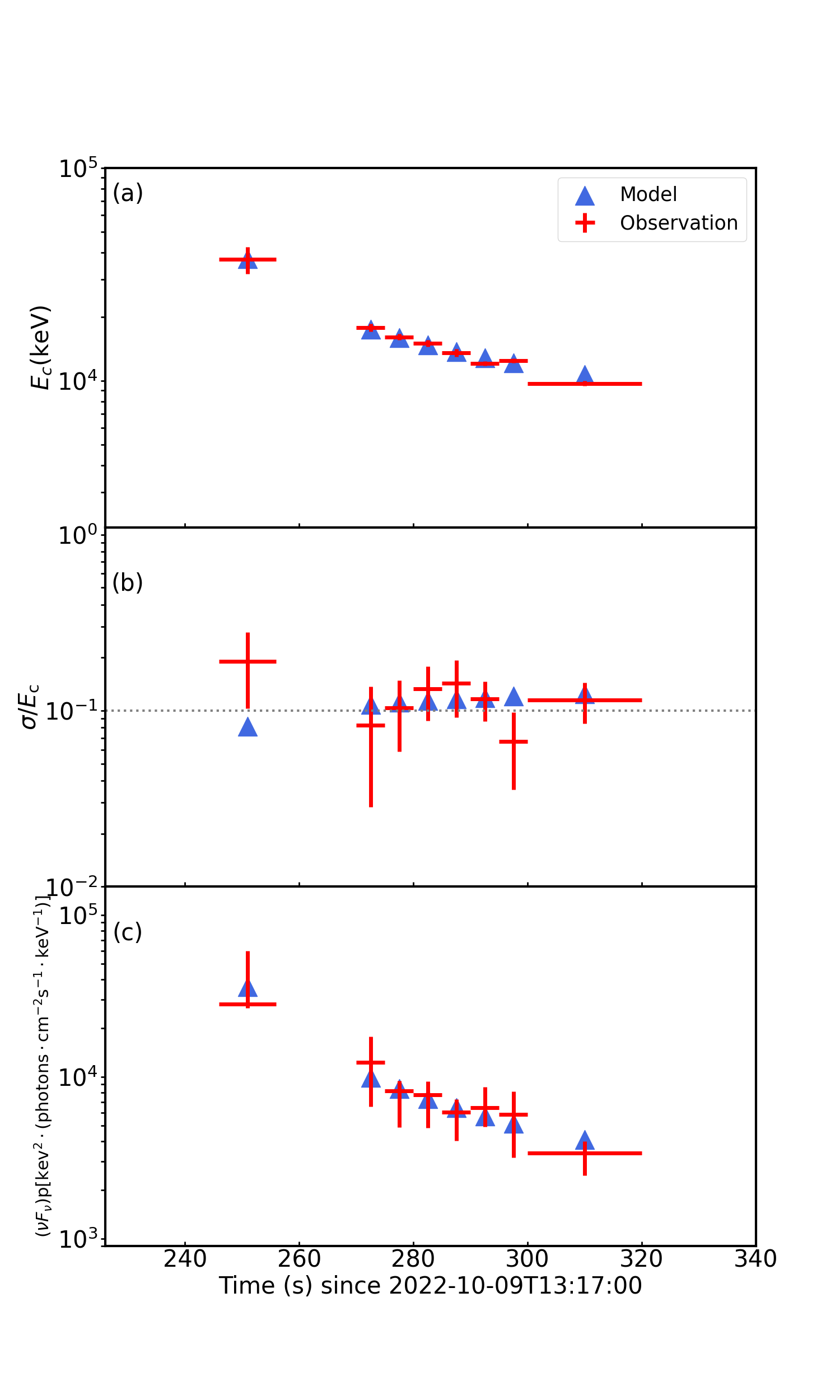}%
\caption{The comparison between the theoretical (blue triangles) and the observed results with 1$\sigma$ error (red crosses). Panel (a) displays the comparison on the time evolution of the central energy $E_{\rm c}$. Panel (b) illustrates the comparison on the time evolution of the width-to-central-energy ratio $\sigma/E_{\rm c}$. Panel (c) depicts the comparison on the time evolution of the peak flux.}
\label{fig:compare2}
 \end{figure}

Furthermore, we transform the theoretical $E'_{\rm c}$ and $\sigma'/E'_{\rm c}$ at the corresponding observed time in the comoving frame to $E_{\rm c}$ and $\sigma/E_{\rm c}$ in the observer's frame. Therefore, we can compare the theoretical values of $E_{\rm c}$ and $\sigma/E_{\rm c}$ with the observed results, which are shown in panels (a) and (b) of Figure \ref{fig:compare2}. Panel (c) in Figure \ref{fig:compare2} further compares the theoretical and observed values of the peak flux of the emission line $(\nu F_{\rm \nu})_{\rm p}$. The red crosses denote
the observed results of the emission line with 1$\sigma$ error, while the blue triangles denote
the theoretical prediction derived by our model. The observed and theoretical results of the emission lines are summarized in Table \ref{table1}. 

Using a power-law fitting to the theoretical evolution of $E_{\rm c}$, we find that $E_{\rm c}$ scales as $(t-t_{\rm 0})^{-1.11}$. Panel (b) shows that our model produces the emission line width-to-central-energy ratio $\sigma/E_{\rm c}$ remaining around 10\%. We also perform a power-law fitting to the theoretical evolution of $(\nu F_{\nu})_{\rm p}$, which yields $(\nu F_{\nu})_{\rm p} \propto (t-t_{\rm 0})^{-1.94}$. The power-law indices characterizing the evolution of both the central energy and the emission line flux align well with the observed values within the error margin. We stress that all the observed evolution behaviors of the emission line can be completely understood by our scenario in a systematic way.

 \begin{table*}
\caption{The first column represents detection epochs. Columns (2)--(4) are the observed results of the emission line, while columns (5)--(7) are the theoretical calculations. The units for the peak flux in column (4) and column (7) are expressed as $ {\rm keV^2 \cdot photons \cdot cm}^{-2}{\rm s}^{-1}\cdot {\rm keV}^{-1}$.}
\label{table1}
\begin{tabular}{lcccccc}  
\hline
\hline
 Time Range  & $E_{\rm c}$(obs)& $\sigma/E_{\rm c}$(obs)  &  $(\nu F_{\rm \nu})_{\rm p}$(obs) & $E_{\rm c}$ & $\sigma/E_{\rm c}$ & $(\nu F_{\rm \nu})_{\rm p}$\\
  (s) & (keV) & & &  (keV) & & \\
 \hline
(246, 256) & $37159\pm 5340$  & $0.19\pm0.09$  & ${28262.84}_{-1629.68}^{+31640.58}$ &   37489.71 & 0.08 & $3.60\times10^4$\\
(270, 275)& $17839\pm760$ &$0.08\pm0.05$  &$ {12312.79}_{-5751.50}^{+5508.77}$& 17508.56 & 0.11& $9.93\times10^3$\\
(275, 280)& $16088\pm564$ & $0.10\pm0.04$& ${8160.14}_{-3292.16}^{+1269.99} $& 15963.80 & 0.11& $8.44\times10^3$ \\
(280, 285)& $15016\pm536$ & $0.13\pm0.05$& ${7745.42}_{-2900.79}^{+1671.66} $& 14722.07 & 0.11& $7.30\times10^3$ \\
(285, 290)& $13552\pm587$ & $0.14\pm0.05$& ${6027.57}_{-2013.45}^{+1269.76}$& 13702.83 & 0.12& $6.42\times10^3$ \\
(290, 295)& $12067\pm311$ & $0.11\pm0.03$& ${6474.13}_{-1547.85}^{+2174.57} $& 12846.45 & 0.12& $5.72\times10^3$ \\
(295, 300)& $12445\pm288$ & $0.07\pm0.03$& ${5877.12}_{-2696.99}^{+2220.94} $& 12112.90 & 0.12& $5.14\times10^3$ \\
(300, 320)& $9726\pm256$ &  $0.11\pm0.03$  & ${3377.77}_{-915.34}^{+604.49}$ & 10666.87 & 0.12& $4.09\times10^3$\\
\hline 
\end{tabular}
\end{table*}

\section{Conclusions} \label{sec:conclusions}
In this work, we comprehensively investigate the effects of the down-Comptonization with jet dynamics on the observed evolution of MeV emission lines in GRB 221009A \citep{2024SCPMA..6789511Z}. We employ a jet dynamical model with an initial Lorentz factor $\Gamma_{\rm 0} =560$ for the jet and a homogeneous circumburst environment characterized by a density of $n_{\rm 0} =0.4\,\rm cm^{-3}$, as constrained by TeV data observed by LHAASO. The temporal evolution of the Lorentz factor is found to deviate from $\Gamma (t) \propto t^{-1}$, instead exhibiting a short plateau phase lasting approximately 10\,s, followed by a gradual decrease before eventually converging into a power-law distribution as $\Gamma(t) \propto t^{-0.42}$. Based on this evolution behavior of the Lorentz factor, we transform the emission line affected by down-Comptonization in the comoving frame to that in the observer's frame. We find that our scenario can successfully reproduce all the observed evolution behaviors of the emission line. The region where down-Comptonization occurs exhibits an approximate density of $2.0\times10^9$\,cm$^{-3}$ and an electron temperature of around 2.0\,keV. This implies the existence of dense and thermal plasma in GRB jet. Our study provides a new direction to the research of extreme cosmic bursts.

\begin{acknowledgments}
We thank the anonymous referee for comments and suggestions that help improve the paper. We appreciate helpful discussions with Zhen Zhang. This work has the financial support of  the National Natural Science Foundation of China (12288102) and the National Key R\&D Program of China (2023YFE0101200 and 2021YFA1600402). J.Y.L. is supported by the Strategic Priority Research Program of Chinese Academy of Sciences, grant No. XDB1160202, the Natural Science Foundation of Yunnan Province (202201AT070158 and 202401AS070045), the Yunnan Revitalization Talent Support Program Young Talent Project, the National Natural Science Foundation of China (12133011), and the International Centre of Supernovae, Yunnan Key Laboratory (202302AN360001). 
J.M. is supported by the National Natural Science Foundation of China (12393813), CSST grant CMS-CSST-2025-A07, and the Yunnan Revitalization Talent Support Program (YunLing Scholar Award). 
S.L.X. and Y.Q.Z. acknowledge the support by the National Natural Science Foundation of China (12273042, 12494572), the Strategic Priority Research Program of the Chinese Academy of Sciences (grant Nos. XDA30050000, XDB0550300) and the National Key R\&D Program of China (2021YFA0718500).  

\end{acknowledgments}
 
\bibliography{grb221009a}{}
\bibliographystyle{aasjournalv7}

\end{document}